\documentclass[letter,twocolumn]{jpsj3}
\usepackage{txfonts}
\usepackage{bm}
\usepackage{color}

\title{Many-Body Chern Numbers of $\nu=1/3$ and $1/2$ States on Various Lattices}

\author{Koji Kudo$^1$, Toshikaze Kariyado$^2$, and Yasuhiro Hatsugai$^{1,3}$}
\inst{$^1$Graduate School of Pure and Applied Sciences, University of Tsukuba, Tsukuba 305-8571, Japan\\
$^2$International Center for Materials Nanoarchitectonics (WPI-MANA), National Institute for Materials Science, 
Tsukuba 305-0044, Japan\\
$^3$Division of Physics, University of Tsukuba, Tsukuba, Ibaraki 305-8571, Japan}
\abst{For various two dimensional lattices such as honeycomb, kagome, and square-octagon, 
the gauge conventions (string gauge) realizing minimum magnetic fluxes that are consistent with 
the lattice periodicity are explicitly given. Then, the many-body interactions of the lattice
fermions are projected into the Hofstadter bands to form pseudopotentials. By using these pseudopotentials, 
the degenerate many-body ground states are numerically obtained. We further formulate a scheme 
to calculate the Chern number of the ground state multiplet by using these pseudopotentials. For 
the filling factor of the lowest Landau level, $\nu=1/3$, a simple scaling form of the 
energy gap is numerially obtained, and the ground state is unique except for the three-fold topological 
degeneracy. This is a quantum liquid, which can be a lattice analogue of the Laughlin 
state. For the $\nu=1/2$ case, the validity of the composite fermion picture is discussed 
in relation to the existence of the Fermi surface. The effects of disorder are also 
described.}

\begin{document}

\maketitle

%\section{INTRODUCTION}
Recent studies have revealed that topology provides a sophisticated view on certain classes of materials. 
The integer quantum Hall (IQH) effect\cite{PhysRevLett.45.494}, which is the quantization of the Hall 
conductance of two-dimensional electrons in strong magnetic field, is explained by
the topological index, i.e., Chern number\cite{PhysRevLett.49.405}. Notably,
intensive studies, conducted in this decade, have revealed that ``a certain class'' is
actually very wide, if the idea of topology is augmented
by the notion of symmetry. Indeed, for noninteracting fermions, an
exploration with various symmetries (and space dimensions) leads to a
``periodic table'' for gapped states containing many
kinds of topologically nontrivial states\cite{PhysRevB.78.195125,PhysRevB.78.195424,doi:10.1063/1.3149495,1367-2630-12-6-065010},
such as a quantum spin Hall state with time reversal symmetry\cite{PhysRevLett.95.226801}. 
%For instance, the Chern number for the IQH
%state takes a nontrivial value only if the time reversal symmetry is
%broken, but there is another class of a topologically nontrivial state
%for two-dimensional electron systems with the time reversal symmetry,
%namely, a quantum spin Hall state. In this way, we can find new 
%topological states. For noninteracting fermions, the exploration of
%various symmetries (and various spatial dimensions) leads to the
%``periodic table'' for topological gapped states, which predicts many
%kinds of topologically nontrivial states. 
 
However, symmetry and dimensionality are not the only directions to
search for novel quantum phases. That is, the incorporation of
electron-electron correlation effects in topological phases is an also 
important issue. The fractional quantum Hall (FQH) effect
\cite{PhysRevLett.48.1559} is a typical example of topologically
nontrivial gapped quantum liquid in which the electron-electron interaction
plays an essential role. The characteristics of the FQH state is well
captured by the wave function proposed by Laughlin\cite{PhysRevLett.50.1395}, and 
the FQH state is highly distinct from that of free fermions\cite{QHE}.
Besides, the composite fermion picture provides a perspective on the correlated electron system\cite{PhysRevLett.63.199}. 
The FQH phase can be specified by the Chern number\cite{PhysRevB.31.3372}, which
involves the twisted boundary condition for its definition;
however, as the
electron-electron interaction is essential, the explicit computation of
the Chern number is not trivial. 

In this letter, several types of lattice models in strong
magnetic field are considered to discuss the electron-electron
interaction effects. 
The physics of lattice models\cite{PhysRevLett.103.105303,PhysRevB.86.165314}, 
related to the FQH system, has been studied in the context of fractional Chern insulators
\cite{PhysRevLett.106.236804,Sheng11,PhysRevX.1.021014,PhysRevB.85.075116},
where the external magnetic field is absent.
Instead, here, we examine lattice models in the external magnetic field. We consider  
six types of lattices: square, Lieb, square-octagon, triangular, honeycomb, and
kagome lattices, and perform numerical analysis on these models. In
order to reduce the computational cost related to the electron-electron
correlation, we project states to the lowest
Landau level (LL), and treat the interaction exactly within the projected
space. It enables us to evaluate the Chern number
explicitly for reasonably large systems. Then, the topological
degeneracy and nonvanishing Chern number are calculated for an electron
filling factor $\nu=1/3$, signaling the FQH states in the lattice models. In addition,
we discuss how the energy spectrum depends on the underlying Bravais lattice.
Furthermore, we also consider $\nu=1/2$ state and discuss their 
consistence with the Fermi liquid states. In the following paragraphs, we first describe
our numerical method, and then explain the details of the
results. 

%\section{PROJECTED POTENTIAL}
Let us begin by introducing the creation-annihilation operators projected
onto a band,
which plays a key role in this paper. We
consider a system with interacting spin-polarized
electrons in a uniform magnetic field on several types of lattices, whose
Hamiltonian is written as $H=H_\text{kin}+H_\text{int}$. The magnetic field is
taken into account by using the Peierls phase in the kinetic term as
$H_\text{kin}=\sum_{\langle i,j\rangle}te^{i\phi_{i,j}}c^{\dagger}_ic_j$,
where $i$ and $j$ are the labels of the sites, $t$ is a hopping parameter, and
$c^{\dagger}_i$ ($c_i$) is the creation (annihilation) operator on site $i$. Note that $\langle i,j\rangle$
indicates the summation over the nearest neighbor pairs of the sites.
Hereafter, we set $t=-1$ for all lattice structures that are considered. The Peierls phase
$\phi_{ij}$ is determined so that an electron traveling around a closed
path acquires a proper phase factor corresponding to the magnetic flux
threading that path.
%The strength of the magnetic field is measured by
%$\phi$, which is the magnetic flux per a unit cell in units of the flux quantum.

\begin{figure}[tb]
  \begin{center}
   \includegraphics[width=\columnwidth]{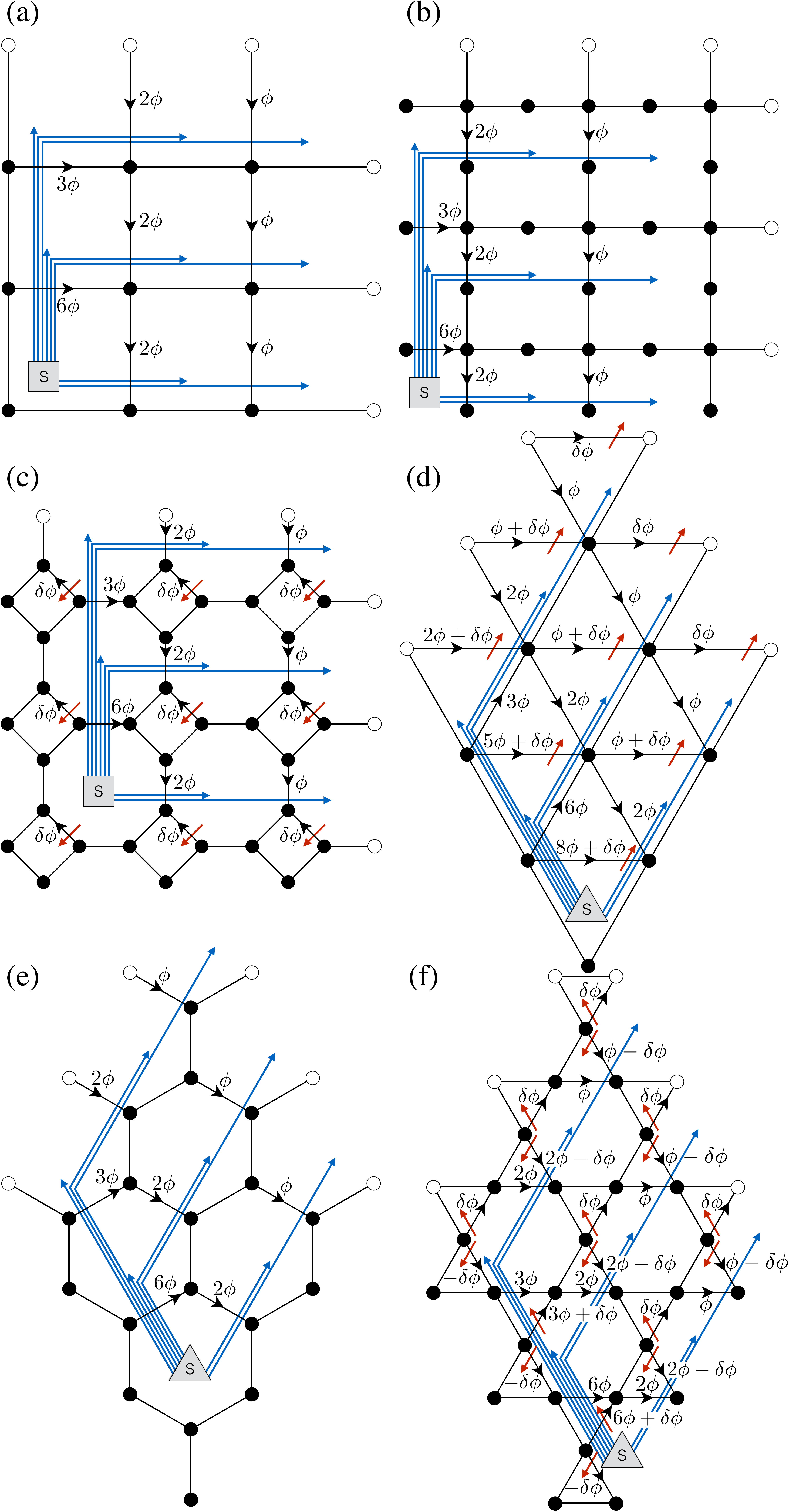}
  \end{center}
  \caption{(Color online) Sketches of $3\times3$ (a) square, (b) Lieb, (c) square-octagon, (d) triangular, (e) honeycomb and (f) kagome lattices with the string gauge under periodic boundary condition.}
% \caption{ The dependence of Energy gap at $\nu=1/3$ and $\nu=1/2$ with $N_\text{e}=5$.
%  \textcolor{red}{The system on }$N\times N$ square lattice (S.L.) and triangular lattice (T.L.) with $\phi=N_{\phi}/N^2$ are considered.}
 \label{fig:fig1}
\end{figure}

In the calculation, the string gauge\cite{PhysRevLett.83.2246} is employed.
Examples of $\phi_{ij}$ assigned by the string gauge for each lattice model
are shown in Fig.~\ref{fig:fig1}. After choosing a unit cell, we set an origin 
$S$ for the strings at an appropriate place in the cell, and draw arrows (strings) to
each unit cell from the origin $S$. To construct a phase $\phi_{ij}=2\pi\phi n_{ij}$,
where $n_{ij}$ is the number of strings that intersect the link $ij$ (the orientation is taken into
account), the strength of the magnetic field per unit cell, except for the one with the
origin $S$, is measured by $\phi$ in units of the flux quantum.
With a uniform magnetic flux, we get
$e^{-i2\pi\phi (N_xN_y-1)}=e^{i2\pi\phi}$ in $N_x\times N_y$ unit cells. ($N_x$ unit cells in one
direction and $N_y$ unit cells in another direction.) It restricts the magnetic flux to
$\phi=N_\phi/(N_xN_y)$ with $N_\phi=1,2,\cdots,N_xN_y$, where $N_\phi$ corresponds to the total magnetic flux.
In the cases of square-octagon, triangular, and kagome lattices
in a uniform magnetic field, it is necessary 
to utilize the strings that transfer the magnetic flux between separated regions in a unit cell, as shown by 
the red arrows in Figs.~\ref{fig:fig1}(c), (d) and (f).
For example, the addition of strings associated with $\delta\phi$ in Fig.~\ref{fig:fig1}(c) realizes a uniform magnetic field
as long as $(\phi-\delta\phi)/S_\text{oc}=\delta\phi/S_\text{sq}$, where $S_\text{oc(sq)}$ is the area of the octagon (square) in the lattice.

For the interaction term, we focus on the nearest neighbor interaction, that is, we use 
$H_\text{int}=\sum_{\langle i,j\rangle}Vc^{\dagger}_ic^{\dagger}_jc_jc_i$,
where $V$ is the strength of the electron-electron interaction. 
In general, it is difficult to solve an interacting
electron problem using full information of the entire Hilbert
space. Therefore, we need to project the
operators into a space spanned by a specific band. 
The lattice model with $\phi\equiv p/q$ ($p,q$: relatively prime)
has $\alpha q$ single-electron bands, where $\alpha$ is the number of sites in a unit cell
with periodic boundary condition. Thus, when the system is put on the $N_x\times N_y$ lattices,
the number of states per band is obtained as
$(\alpha N_xN_y)/(\alpha q)=N_xN_y/q$. For $p\ll q$, the energies between the lowest and the
$p$-th bands form the LL in the large $q$ limit.
Therefore ``the lowest Landau level'' is defined as a group of these states, and we 
focus on the projection to this LL with the Landau degeneracy $(N_xN_y/q)\times p=N_{\phi}$.

%The ``band'' is defined
%as a group of eigenstates of $H_{\text{kin}}$ (not $H$) whose
%eigenvalues are separated from those of the other eigenstates.
%In strong magnetic field, a LL forms a band in this sense, and in this
%paper, we focus on the projection to the lowest LL
%with the Landau degeneracy $N_{\phi}=\phi N_x N_y$, though the
%following arguments could be applied to different situations.

A multiplet is numerically constructed using the eigenvectors belonging to the lowest LL
as $\psi=(\psi_1,\psi_2,\cdots,\psi_{N_{\phi}})$, and the projected creation operator is
defined as $\tilde{c}^{\dagger}_i=(\bm{c}^{\dagger}P)_i$, where
$P=\psi\psi^{\dagger}$, $\bm{c}^{\dagger}=(c^{\dagger}_1,c^{\dagger}_2,\cdots,c^{\dagger}_{N_\text{site}})$ and
$N_\text{site}=\alpha N_xN_y$ \cite{PhysRevB.86.205424,1367-2630-15-3-035023}.
By using these projected creation operators, the Hamiltonian is projected into the lowest LL by replacing
$c^{\dagger}_i$, $c_i$ with $\tilde{c}^{\dagger}_i$,
$\tilde{c}_i$. Since we have
$\{\tilde{c}^{\dagger}_i, \tilde{c}_j\}\neq \delta_{ij}$ and
$\{\tilde{c}^{\dagger}_i, \tilde{c}^{\dagger}_j\}=\{\tilde{c}_i, \tilde{c}_j\}=0$,
the canonical anticommutation relations are no longer satisfied,
and therefore, the ordering of fermions is important. The Hamiltonian is used 
in the form of a semi-positive definite as
\begin{align}
  \tilde{H}_\text{int}
  =\sum_{\langle i,j\rangle}V\tilde{c}^{\dagger}_i\tilde{c}^{\dagger}_j\tilde{c}_j\tilde{c}_i
  =\sum_{k,l,m,n}V_{klmn}d^{\dagger}_kd^{\dagger}_ld_md_n.
\end{align}
Here, $V_{klmn}=\sum_{<i,j>}V(\psi_k)_i^{\ast}(\psi_l)_j^{\ast}(\psi_m)_j(\psi_n)_i$, 
the summation over $k,l,m,n$ is restricted to the states
on the lowest LL, and $d^{\dagger}_k$ is the creation operator of the state $k$ as 
$d^{\dagger}_k=\bm{c}^{\dagger}\psi_k$. 
Here, we choose $V$ such that the typical energy
scale of the electron-electron interaction is much larger than the
energy width of the lowest LL, and consider only the interaction term. To
diagonalize $\tilde{H}_\text{int}$ for the many-electron states, we need 
the matrix element using
$|\Psi_i\rangle=d_{i_1}^{\dagger}\cdots d_{i_{N_\text{e}}}^{\dagger}|0\rangle$
as the basis for the $N_\text{e}$-electron system.

%\section{EXACT DIAGONALIZATION}
%\subsection{MANY-BODY GAP}
We first calculate the energy spectra at the LL filling $\nu=N_\text{e}/N_\phi=1/3$ and $1/2$, especially focusing
on the gap above a ground state multiplet. Here, 
%the ground state multiplet, the ground state with multiple degeneracy, is defined as a
%group of states whose energy are close to the lowest energy, and the
%energy variation within the group is much smaller than the gap between
%the other excited states and the ground state.
if $m$ is the minimum integer satisfying
$(\tilde{E}_{m+1}-\tilde{E}_1)/(V\rho E_\text{LG})>10^{-3}$, where $\tilde{E}_{i}$ is the $i$-th eigenvalue
of $\tilde{H}_\text{int}$, $E_\text{LG}$ is the Landau gap of the non-interacting case and $\rho=N_\text{e}/N_\text{site}$, 
we define the first $m$ states as the $m$-fold degenerate ground states.
%Figure~\ref{fig:fig1}(a) compares the energy gaps above the ground state
%multiplet for six types of the lattice structures: square, Lieb,
%square-octagon, triangular, honeycomb, and kagome, where the first three
%have the square Bravais lattice while the last
%three have the hexagonal one. In Fig.~\ref{fig:fig1}(a), the values of
%energy gap $\Delta\tilde{E}$ are scaled by $V\rho E_\text{LG}$, where
%$\rho=N_\text{e}/N_\text{site}$ and $E_\text{LG}$ is the Landau gap that 
%is defined as the difference between the lowest energy on the lowest
%LL and the second LL. After scaling, it is
%revealed that the numerically obtained energy gaps for the different types
%of lattices have the similar values as far as the underlying Bravais
%lattice is the same. 

\begin{figure}[t]
  \begin{center}
   \includegraphics[width=\columnwidth]{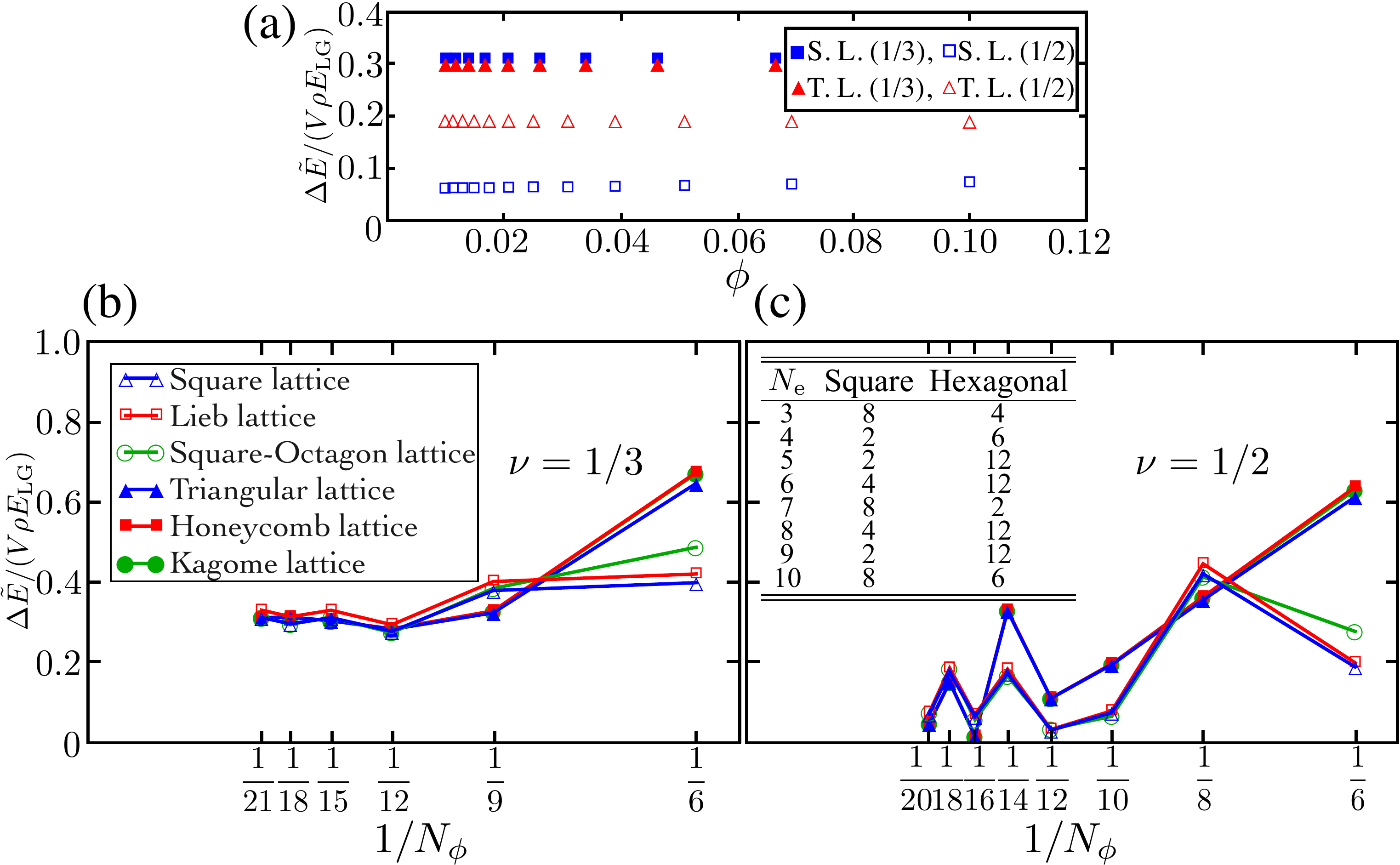}
   \caption{(Color online) Scaled energy gaps as functions of (a) the magnetic flux $\phi$ and (b,c) the inverse of Landau degeneracy $1/N_\phi$.
   The results are displayed for 5 electrons on square lattices (S.L.) and triangular lattices (T.L.) at $\nu=1/3$ and $1/2$ in (a). The table in (c) represents
   the degeneracies of the ground states at $\nu=1/2$ for the square Bravais lattice and the hexagonal one.}
%   (a) The system on $N\times N$ square lattice (S.L.) and triangular lattice (T.L.) with $\phi=N_{\phi}/N^2$ at $\nu=1/3$ and $1/2$ are considered.
%   The number of electrons is $N\text{e}=5$. 
%   (b,c) The systems on the $N_phi\times N_\phi$ various lattices with $\phi=1/N_\phi$ are considered.
%   The table in (b) represents the degeneracies of the ground states at $\nu=1/2$ for the lattice models
%   which have the square Bravais lattice (square, Lieb, square-octagon lattices) and the hexagonal one (triangular, honeycomb, kagome lattices).}}
%   \caption{The scaled energy gap at (a)\,$\nu=1/3$ and (b)\,$\nu=1/2$ on \textcolor{red}{various} lattice models. The systems on the $N_\phi\times N_\phi$
%   lattices with $\phi=1/N_\phi$ are considered. The table in (b) represents the degeneracies of the ground states at $\nu=1/2$ 
%   \textcolor{red}{for the lattice models which have the square Bravais lattice (square, Lieb, square-octagon lattice) and the hexagonal one (triangular, honeycomb,
%   kagome lattices).}}
    \label{fig:fig2}
  \end{center}
\end{figure}

In Fig.~\ref{fig:fig2}(a), the energy gaps of the systems with
$N\times N$ square and triangular lattices are plotted as functions of
$\phi=N_{\phi}/N^2$.
Since the energy scale is described by the Landau gap
of the non-interacting case $E_\text{LG}$ and
$\sum_{<i,j>}n_in_j\sim\sum_{<i,j>}\rho^2\sim\rho$, 
the energy gap $\Delta\tilde{E}$ is scaled by $V\rho E_\text{LG}$.
%, where $\rho=N_\text{e}/N_\text{site}$ and $E_\text{LG}$ is the Landau gap 
%of the non-interacting case.
%that is defined as the difference between the lowest energy on the lowest
%LL and the second LL.
The results in Fig.~\ref{fig:fig2}(a)
show that the scaling law $\Delta\tilde{E}\propto V\rho E_\text{LG}$
is valid in the wide range of $\phi$, regardless of the lattice types.
Since the Landau gap $E_\text{LG}$ and $\rho=(N_\text{e}/\alpha N_{\phi})\phi$ are
proportional to $\phi$, Fig.~\ref{fig:fig2}(a) indicates the relation
$\Delta\tilde{E}\,\propto\phi^2\,\propto\rho^2$, which 
means that the excitations are local at both $\nu=1/3$ and
$1/2$.
%I COULD NOT SEE THE POINT OF THIS STATEMENT. $\phi$ IS THE
%STRENGTH OF THE MAGNETIC FILED, RATHER THAN THE SYSTEM SIZE. 
%}

The difference between $\nu=1/3$ and $1/2$ becomes clear when we consider
the dependence of energy gaps on the Landau degeneracy $N_\phi$. 
The numerically obtained $1/N_\phi$ dependence of the energy gaps
%for the square and the triangular lattices
is shown in Figs.~\ref{fig:fig2}(b) and (c),
where we consider six types of lattice structures: square, Lieb,
square-octagon, triangular, honeycomb, and kagome. The first three
have square Bravais lattice while the last
three have hexagonal Bravais lattice.
Here, the systems with $\phi=1/N_{\phi}$ on the
$N_{\phi}\times N_{\phi}$ lattices are considered. Note that the scaling
found in Fig.~\ref{fig:fig2}(a) is independent of $\phi$.
In Figs.~\ref{fig:fig2}(b) and (c), the energy gaps behave similarly as a 
function of $1/N_\phi$ if the underlying Bravais
lattice is the same. In addition, Fig.~\ref{fig:fig2}(b) indicates
the finite energy gap in the large $N_\phi$ limit, which is
consistent with the Laughlin state as a ground state for $\nu=1/3$. 
On the other hand, for $\nu=1/2$, their 
$1/N_\phi$ dependence is clearly different from that for $\nu=1/3$ and 
can be consistent with the gap closing behavior.

%\subsection{TOPOLOGICAL DEGENERACY}
Another important quantity that characterizes the difference between the
odd-denominator filling fractions (e.g. $\nu=1/3$) and the even-denominator
ones (e.g. $\nu=1/2$) is the degeneracy of the ground state.
The ground states for $\nu=1/3$ are always
accompanied by the three-fold topological degeneracy
\cite{PhysRevLett.55.2095}. This feature holds irrespective of the
lattice type, which is explained by the translation of the
center-of-mass.

In contrast, the degeneracy of the ground states at $\nu=1/2$ has no
such universal feature. The ground state degeneracy for $\nu=1/2$ is
shown in the inset table in Fig.~\ref{fig:fig2}(c).
The degeneracy is always even, which is supported by the center-of-mass translation,
%The residual degeneracy
%is caused by the relative translation
%There exist symmetries other than the center-of-mass translation symmetry \cite{PhysRevLett.55.2095}, and 
and depends on the number
of electrons and the underlying Bravais lattice.
%rotational symmetry of the Bravais lattices. 
The many-electron system having interactions in a magnetic
field with $\nu=1/2$ is mapped to the Fermi liquid with composite
fermions \cite{PhysRevLett.63.199,PhysRevB.47.7312}. 
%If there is $Q$-fold rotational symmetry in the underlying Bravais lattice on the system without magnetic field,
%the band structure has the same symmetry around the origin in the reciprocal space. 
Without any magnetic field, there is a $Q$-fold rotational symmetry around the origin in the band structure, when 
the considered lattice type has the square Bravais lattice ($Q$=4) or the hexagonal one ($Q$=6).
Thus, as long as $N_1=N_2$ and the system
is not too small, the ground state of the many-electron state forms a
close shell and the total momentum is zero, 
when the number of electrons is $1+Qn$, ($n$: integer).
In fact, in the table in Fig. \ref{fig:fig2}(c), the ground states have
no degeneracy for $N_\text{e}=1+Qn$, if we ignore the factor of two given by the
center-of-mass translational symmetry ($N_\text{e}=$5 and 9 for the square Bravais
lattice and $N_\text{e}=7$ for the hexagonal one).
Besides, the trend seen in Fig. \ref{fig:fig2}(c) is that the energy gaps of the ground state forming 
a close shell are larger than those of the other states. 
These observations are consistent with the existence of the Fermi surface of the composite fermions.
%Besides, this picture is considered to be reflected in some peaks not seen in Fig. \ref{fig:fig2}(b) but (c).
%Especially, a trend is seen that the energy gaps of the ground state forming a close shell are larger than those around them,
%and it also suggests that the excitation mechanism at $\nu=1/2$ is different from that at $\nu=1/3$.

%\section{HALL CONDUCTANCE}
As we have seen, the ground state at $\nu=1/3$ has a three-fold
topological degeneracy. Then according to the Niu-Thouless-Wu formula \cite{PhysRevB.31.3372}, 
the Hall conductance is given by $\sigma_{xy}=\frac{e^2}{h}\frac{C}{m}$,
where $C=\frac{1}{2\pi i}\int_{T^2}\hbox{Tr }{\bm F}$, ${\bm F}=d{\bm A}+{\bm A}^2$,
and ${\bm A}$ is the non-Abelian Berry connection\cite{Berry84}, which is given by
the ground state multiplet
$\Phi=(|G_1\rangle,\ldots,|G_m\rangle)$ as ${\bm A}=\Phi^{\dagger}d\Phi$
\cite{doi:10.1143/JPSJ.73.2604,doi:10.1143/JPSJ.74.1374}. Here, $|G_i\rangle$ are the ground
states with $m$-fold topological degeneracy ($\langle G_i|G_j\rangle=\delta_{ij}$). 
The domain of integration $T^2$ is a parameter space given by the
twisted boundary condition. We evaluate the Hall conductance by
computing the Chern number explicitly using the ground state multiplet. 

%\subsection{METHOD}
To obtain the Chern number, we impose a twisted boundary condition 
as $c^{\dagger}_{(n_x+N_x,n_y,s)}=e^{i\theta_x}c^{\dagger}_{(n_x,n_y,s)}$ and 
$c^{\dagger}_{(n_x,n_y+N_y,s)}=e^{i\theta_y}c^{\dagger}_{(n_x,n_y,s)}$, where $(n_x,n_y,s)$ is the
site index ($n_x\in\{1,\cdots,N_x\}$, $n_y\in\{1,\cdots,N_y\}$, $s\in\{1,\cdots,\alpha\}$).
The eigenvectors of the lowest LL $\psi_k$'s depend on
$\theta=(\theta_x,\theta_y)$ through the dependence of $H_{\text{kin}}(\theta)$,
which causes a modification on the projected interaction
Hamiltonian as 
\begin{align}
  \tilde{H}_\text{int}(\theta)
  =\sum_{k,l,m,n}V_{klmn}(\theta)d^{\dagger}_k(\theta)d^{\dagger}_l(\theta)d_m(\theta)d_n(\theta),
\end{align}
where $d^{\dagger}_k(\theta)=\bm{c}^{\dagger}\psi_k(\theta)$.

By diagonalizing $\tilde{H}_{\text{int}(\theta)}$, we obtain an $m$-component
ground state multiplet as $\Phi(\theta)=(|G_1(\theta)\rangle,|G_2(\theta)\rangle,\ldots,|G_m(\theta)\rangle)$,
where $|G_i(\theta)\rangle$'s are the $m$-fold degenerate ground states
of $\tilde{H}_\text{int}$ satisfying $\langle G_i|G_j\rangle=\delta_{ij}$.
By using this multiplet, the Chern number is evaluated by applying the
method proposed in ref.~\citen{doi:10.1143/JPSJ.74.1674}.
A $U(1)$ link variable on a discretized link in the parameter space is defined as
$U_{\mu}(\theta_l)=\frac{1}{N_{\mu}(\theta_l)}\,\det[\Phi^{\dagger}(\theta_l)\Phi(\theta_l+\Delta_{\mu})]$,
where $\Delta_{\mu}$ represents the displacement in the
direction $\mu=x,y$ at $\theta_l$ and $N_{\mu}(\theta_l)=|\det[\Phi^{\dagger}(\theta_l)\Phi(\theta_l+\Delta_{\mu})]|$.
As seen from the definition, the link variables require the
computation of the overlap between the ground states at $\theta_l$ and
$\theta_l+\Delta_\mu$.

When $\tilde{H}_\text{int}(\theta)$ is diagonalized by the orthonormal basis
$\Psi(\theta)=(|\Psi_1(\theta)\rangle,|\Psi_2(\theta)\rangle,\cdots,|\Psi_{N_\text{D}}(\theta)\rangle)$
($N_\text{D}={}_{N_{\phi}}\!C_{N_\text{e}}$),
the eigenvalue equation $\tilde{h}_\text{int}(\theta){\bm u}_i(\theta)=\tilde{E}_i(\theta){\bm u}_i(\theta)$,
where $\tilde{h}_\text{int}=\Psi(\theta)^{\dagger}\tilde{H}_\text{int}(\theta)\Psi(\theta)$, is given and the ground state is expressed as
\begin{align}
  |G_k(\theta)\rangle=\Psi(\theta){\bm u}_k(\theta),
\end{align}
where $\tilde{E}_k(\theta)$ is one of the energies of the ground state multiplet.
Using this expression,
the overlap between the states with different boundary conditions,
$\theta$ and $\theta'$, is given by
\begin{gather}
  \langle G_k(\theta)| G_l(\theta')\rangle={\bm u}_{G_k}^{\dagger}(\theta)O(\theta,\theta'){\bm u}_{G_l}(\theta'),\\
  O(\theta,\theta')=\Psi^{\dagger}(\theta)\Psi(\theta').
\end{gather}
The $(i,j)$ element of $O(\theta,\theta')$ is expressed as 
$O_{ij}(\theta,\theta')=\det[\tilde{\psi}_i(\theta)^{\dagger}\tilde{\psi}_j(\theta')]$,
where $\tilde{\psi_i}(\theta)=(\psi_{i_1}(\theta),\cdots,\psi_{i_{N_\text{e}}}(\theta))$.~\cite{detail1}

After obtaining the link variable in the above way, the lattice Berry curvature is defined as
\begin{align}
  \tilde{F}_{12}(\theta_l)=\mathrm{Log}[U_1(\theta_l)U_2(\theta_l+\Delta_1)U_1^{-1}(\theta_l+\Delta_2)U_2^{-1}(\theta_l)]
\end{align}
and $-\pi<\tilde{F}_{12}(\theta_l)/i\leq\pi$. The function $\mathrm{Log}$ means taking the principle branch of the
logarithm. By definition, $\tilde{F}_{12}(\theta_l)$ is invariant under the $U(m)$ gauge transformation
$\Phi(\theta)\rightarrow\Phi'(\theta)=\Phi(\theta)\omega(\theta)$. Now, the Chern number on the lattice is given as
\begin{align}
  \tilde{C}=\frac{1}{2\pi i}\sum_l\tilde{F}_{12}(\theta_l),
\end{align}
where the summation is taken over all the mesh points in the
parameter space. It is guaranteed that $\tilde{C}$ is always integral
and becomes exact in the limit of the fine mesh. 

\begin{figure}[t]
  \begin{center}
   \includegraphics[width=\columnwidth]{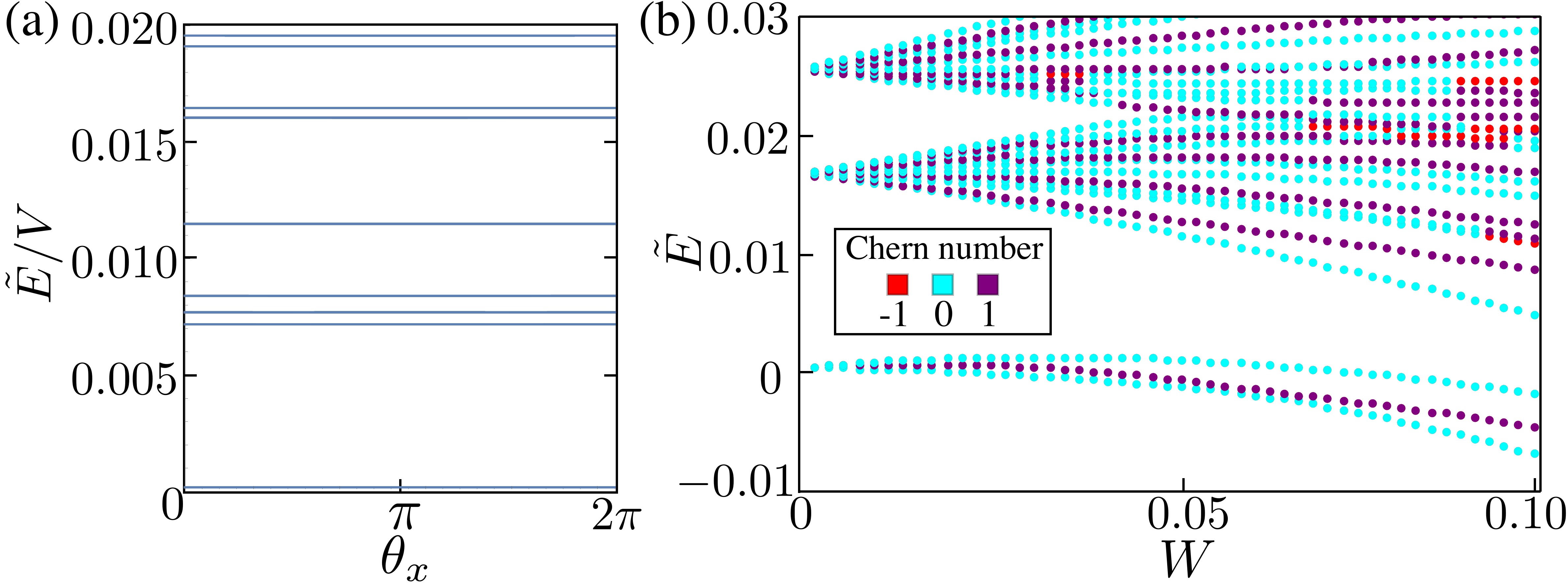}
   \caption{(Color online) (a)The eigenvalue of $\tilde{H}_\text{int}$ at $\nu=1/3$ against $\theta_x$ for $\theta_y=0$.
   The system with $12\times12$ square lattices for $\phi=1/12$ is considered.
   The ground state is accompanied with the $3$-fold degeneracy in arbitrary parameters $\theta$ and well separated from the first excited states.
   (b)The eigenvalue of $\tilde{H}_\text{int}+\tilde{H}_\text{imp}$ at $\nu=1/3$ under the periodic boundary condition against the strength of random potential $W$.
   The system with $9\times9$ square lattices
   for $\phi=1/9$ is considered and we set $V=1$. The Chern numbers are expressed by the color of plots.}
   \label{fig:fig3}
  \end{center}
\end{figure}

%\subsection{RESULT}
We diagonalize $\tilde{h}_\text{int}(\theta)$ for $\nu=1/3$ and its energy
$\tilde{E}_i(\theta)$ is plotted as shown in Fig.~\ref{fig:fig3}(a). There is no
level crossing between the ground state multiplet with the three-fold topological degeneracy and excited states.
For $\nu=1/3$, the Chern number of the ground state multiplet is
1. This means that the 
quantized Hall conductance is $e^2/3h$.
On the other hand, the other excited states have
$3n$-fold $(n:\text{integer})$ degeneracy and the Chern number is $n$,
which indicates that the average of the Hall conductance 
is written as $\langle\sigma_{xy}\rangle=e^2/3h$ for any temperature.
%\textcolor{red}{which is consistent with the case of the high-temperature limit 
%where the classical picture is applied. 
At $\nu=1/2$, as mentioned previously, 
the degeneracy of the ground states is generically
$2n$ $(n:\text{integer})$. In this case, the Chern number of
the ground state multiplet is $n$. In general, the Hall
conductance specified by the Chern number of the ground state multiplet
for the filling factor $\nu$ is evaluated as $\nu e^2/h$.

We also investigate the effects of disorder. We limit ourselves to
the cases where the disorder potential is sufficiently small compared with
the Landau gap, which allows us to discuss the impurity effects within
the states projected to the lowest LL. We define the projected impurity potential as
\begin{align}
 \tilde{H}_\text{imp}
 =\sum_{i}w_i\tilde{c}^{\dagger}_i\tilde{c}_i,
\end{align}
where $w_i=Wf_i$ is the site potential at a site $i$, $f_i$ represents
uniform random numbers between $[-1/2,1/2]$, and $W$ is the strength of
the random potential.
%\begin{figure}[t!!]
%  \begin{center}
%   \includegraphics[width=\columnwidth]{fig_latticeFQHE4/fig5.pdf}
%   \caption{The energy at $\nu=1/3$ under the periodic boundary condition against the strength of random potential $W$. The system with $9\times9$ square lattices
%   for $\phi=1/9$ is considered. The Chern numbers are expressed by the color of plots.}
%    \label{fig:fig5}
%  \end{center}
 %\end{figure}
In Fig. \ref{fig:fig3}(b), the energy spectrum of $\tilde{H}_\text{int}+\tilde{H}_\text{imp}$
(for $\theta=0$) is
plotted against $W$ with the Chern number indicated using different colors.
In general, the topological degeneracy is lifted by the disorder in any value of
$\theta$, and therefore, the Chern numbers can be individually
assigned to each lifted state \cite{PhysRevB.40.12034}. More
specifically, the three-component ground state multiplet is split into
three states, where one state carries a Chern number of 1, while the
other two carry 0. This is topological stability.
Furthermore, the numerical results suggest that the
state with the lowest energy is always trivial in terms of the Chern number,
which implies that the Hall conductance is zero when the temperature is
smaller than the small energy gap within the lifted ground state
multiplet. 

%\section{SUMMARY}
To summarize, we construct the Peierls phase by using the string gauge for various types of
lattices and analyze the many-electron states by using the Hamiltonian projected to the lowest LL.
By diagonalizing the pseudopotential, a simple scaling form of the energy gap is obtained.
The results for $\nu=1/3$ indicate that the ground states accompanied by three-fold topological 
degeneracy are consistent with the Laughlin state. On the other hand, the degeneracy of the ground state
for $\nu=1/2$ depends on the type of lattice structure, which is discussed in terms of the composite
fermion picture using the existence of the Fermi surface. We further formulate a method
to compute the Chern number of the ground state multiplet using the pseudopotential.
This method is applied to the lattice analogue of the Laughlin state and the effects of disorder 
are discussed with the Chern number.
\begin{acknowledgment}
This work is partly supported by Grants-in-Aid for Scientific Research, (KAKENHI), Grant numbers 17H06138, 16K13845 and 25107005.
\end{acknowledgment}
% \begin{thebibliography}{9}
%  \bibitem{jpsj} The abbreviation for JPSJ must be ``J. Phys. Soc. Jpn.'' \note{in the reference list}.
%  \bibitem{instructions} More abbreviations of journal titles are listed in ``Instructions for Preparation of Manuscript''.
%  \bibitem{etal} The use of $B!H(Bet al.$B!I(B is not accepted in principle, therefore, all the authors must be listed.
%  \bibitem{ibid} The term $B!H(Bibid.$B!I(B should not be used even if the same journal or book is cited with different page numbers.
%  \bibitem{Errata} Errata should be listed under the same reference number.
% \end{thebibliography}
%\nocite{*}
\bibliographystyle{jpsj}
\bibliography{citation}
\end{document}